\documentclass[10pt,a4paper,reqno,oneside]{article}
\usepackage{amsmath, amsthm, amsbsy} 
\usepackage{pstricks}
\usepackage{algorithmic,algorithm}
\usepackage{graphicx}
\usepackage{url}
\usepackage{booktabs}
\usepackage{a4,amssymb,verbatim}
\usepackage{setspace}
\usepackage{caption,bm,subcaption}
\usepackage[authoryear]{natbib}
\newcommand{\ba}{\begin{eqnarray*}}
\newcommand{\ea}{\end{eqnarray*}}

\newcommand{\bq}{\begin{eqnarray}}
\newcommand{\eq}{\end{eqnarray}}
\newcommand{\mse}{\operatorname{MSE}}

\DeclareMathOperator*{\argmax}{argmax} 

\long\def\symbolfootnote[#1]#2{\begingroup%
\def\thefootnote{\fnsymbol{footnote}}\footnote[#1]{#2}\endgroup} 

\newcommand{\blind}{1}

\begin{document}

\def\spacingset#1{\renewcommand{\baselinestretch}%
{#1}\small\normalsize} \spacingset{1}


\if1\blind { \title{\bf Gaussian process modeling of 
    heterogeneity and discontinuities using Voronoi tessellations}
  \author{Christopher A. Pope\thanks{
      The first author gratefully acknowledges the financial support of the John E Crowther -- Martin Clarke Research Foundation.}, John Paul Gosling, Stuart Barber,\hspace{.2cm}\\
    School of Mathematics, University of Leeds,\\
    Jill Johnson, \\
    School of Earth and Environment, University of Leeds,\\
    Takanobu Yamaguchi, Graham Feingold,\\
    Chemical Sciences Division, Earth System Research Laboratory,\\
    National Ocean and Atmospheric Administration, Boulder, CO 80305\\
    and\\
    Paul G. Blackwell, \\
    School of Mathematics and Statistics, University of Sheffield.}
  \maketitle
} \fi

\if0\blind
{
  \bigskip
  \bigskip
  \bigskip
  \begin{center}
    {\LARGE\bf  Gaussian process modelling of 
    heterogeneity and discontinuities using Voronoi tessellations}
\end{center}
  \medskip
} \fi

\bigskip
\begin{abstract}
  Many methods for modelling functions over high-dimensional spaces assume global
  smoothness properties; such assumptions are often violated in
  practice.  We introduce a method for modelling functions
  that display heterogeneity or contain discontinuities. The
  heterogeneity is dealt with by using a combination of Voronoi
  tessellation, to partition the input space, and separate Gaussian
  processes to model the function over different regions of the partitioned
  space. The proposed method is highly flexible since it allows the
  Voronoi cells to combine to form regions, which enables non-convex
  and disconnected regions to be considered.  In such problems,
  identifying the borders between regions is often of great importance
  and we propose an adaptive sampling method to gain extra information
  along such borders.  The method is illustrated by simulated examples
  and an application to real data, in which we see improvements in
  prediction error over the commonly used stationary Gaussian process
  and other non-stationary variations. In our application, a
  computationally expensive computer model that simulates the
  formation of clouds is investigated, the proposed method more accurately predicts
  the underlying process at unobserved locations than existing
  emulation methods.
\end{abstract}

\noindent%
{\it Keywords:}  emulation, Gaussian process, locating discontinuities, Voronoi tessellation.  
\vfill

\newpage
\spacingset{1.45} 
\section{INTRODUCTION}
\noindent

Many methods used to model uncertainty about functions make assumptions about the smoothness of the function over its input space.  By smoothness, we are referring to the assumption that minor perturbations in the inputs lead to only minor changes in the function's output.  However, there are many examples of small fluctuations in the input causing great changes in a function's output. We discuss one such example in Section~\ref{EG}, where small changes in the input parameters of a computer model of a cloud formation process lead to large changes in the properties of the process. Naively modelling these functions using methods that rely on the assumption of smoothness can lead to poor results when the models are used for analysis such as prediction \citep{paciorek1}. Methods which make fewer smoothness assumptions, such as mixtures of thin plate splines \citep{wood1}, local linear regression \citep{cleveland1} and wavelet-based imputation \citep{Heaton1}, have been used to build approximations for the underlying functions of these processes. By making so few assumptions about the data, methods such as these have the drawback that a large number of observations are needed to build an accurate model of the underlying process. We propose an approach where the space is divided into disjoint regions such that the process can be assumed to be smooth within each region, while allowing for abrupt changes at the boundaries of the regions.  Within each region, the local smoothness assumption means that the model can function with relatively few data points.  The method in this paper is applicable to many situations where a process displays heterogeneity or discontinuities separating regions that are internally smooth.

One well-established method for spatial modelling is Gaussian process regression or kriging \citep{cressie1, handcock1993}. By using a Gaussian process to model the underlying function, we are making an assumption of smoothness in the underlying function over the entire input space. As mentioned previously, this assumption is rarely justified. To deal with this, adaptations to the stationary Gaussian process methodology must be made to accommodate non-stationarity. Two of the main methods that have been focused on in the literature are changes to the covariance function, such as spatial deformations \citep{sampson2,Schmidt1} or convolution based methods \citep{higdon1,risser2015}, and the use of independent Gaussian processes over a partitioned space, such as treed Gaussian process \citep[TGP;][]{gramacy1} or piecewise Gaussian process \citep[PGP;][]{kim1}.  This paper's focus is the latter of the two categories, and readers interested in adaptations to the covariance function are directed to \citet{Risser1} for a review. 

In order to fit a piecewise Gaussian process model, we must specify a technique for partitioning the input space. Both of the methods mentioned previously partition the input space to give regions with straight boundaries: treed partitioning does so using non-overlapping lines parallel to the input axes and Voronoi tessellation uses the Euclidean distance from a set of centres to create Voronoi cells. In this paper, we shall focus on partitioning the input space using Voronoi tessellation due to its flexibility compared to treed partitioning. We allow Voronoi cells to join together to create larger, more flexible, joint regions. Once we have specified the partition of the input space, we can fit separate Gaussian processes to each region. The models which are built using Voronoi tessellations in \citet{kim1} are a special case of this where there is no combining of cells and severe constraints are applied to the locations of the centres. The joining of  tessellation cells allows more complex regions, such as when one region is surrounded by another, or non-convex shaped regions, without the loss of information that is intrinsic to building regions that can only be a single independent cell. Very importantly, we also look to allow a greater range of models than the PGP model of \citet{kim1} by changing the prior distribution of the centres that defines the cells of the tessellation. 

We also tackle the problem of adaptive sampling where the aim is to locate the discontinuities in the function and major changes in the function's response to inputs. This gives us a means of reducing uncertainty about the regions' boundaries via further sampling. Traditional sampling methods are not geared towards this objective and are shown by examples to perform worse than the proposed method in the presence of different regions.  

Our application, the System for Atmospheric Modelling (SAM) model \citep{Khairoutdinov1} is very computationally expensive, a single run taking over ten hours, and so the model behaviour over the six-dimensional parameter space cannot be fully explored using it directly. Hence, the proposed approach to generate a statistical representation of the model provides a means by which the cloud behaviour can be more rigorously examined. The clouds that the SAM model simulates are particularly sensitive to aerosol concentrations in the atmosphere and meteorological conditions, where small changes in temperature and humidity profiles can  impact strongly on whether clouds form or not, and how thick/reflective they are. The proposed statistical method, which does not assume global smoothness, allows us to represent and explore the SAM model more accurately than existing methods.

We give a brief overview of the stationary Gaussian process model in Section~\ref{GP}. In Section~\ref{PGP}, we describe the proposed partitioning technique and the MCMC methods needed to remove dependence on the partition structure for inference. Section~\ref{SM} introduces an adaptive sampling method used to better define the location of a discontinuity. Section~\ref{TOY} demonstrates the technique on a simulated example, and Section~\ref{EG} shows the method applied to a real example. We finish with a discussion of the method and possible future extensions in Section~\ref{DIS}.\

\section{MODELLING USING A GAUSSIAN PROCESS}\label{GP}

We consider a measured attribute $y\in \mathbb{R}$ corresponding to inputs $\mathbf{x}\in \mathcal{X} \subseteq\mathbb{R}^d$. In a spatial setting, $\mathbf{x}$  corresponds to the spatial location; however, we are not restricted to this, and more general high-dimensional inputs can be used for other applications in analysing computer model output. We represent the relationship between the input and output by a function: $y=\eta(\mathbf{x})$.  The output $y$ is not necessarily a scalar, though we only consider the case of a scalar output in this paper. Examples of multivariate outputs are given by \citet{rougier2}, \citet{fricker1} and \citet{conti1}; the proposed  approach could be extended for multivariate outputs.  Spatial process data are typically measured with natural variation or error, so repeated observations of identical $\mathbf{x}$ results in different outputs $y$. The approach we propose is also applicable to cases in which the output is deterministic, where multiple applications of the same $\mathbf{x}$ will result in the same output $y$, which is often found when considering computer model output, such as those seen in \citet{sacks1}, \citet{linkletter1} and \citet{currin1}. In this paper, we employ the Gaussian process as a regression model; however, this is just one possible choice and other statistical models could be used.

The conditional mean of $\eta(\mathbf{x})$ given a vector of coefficients $\boldsymbol{\beta}$ is given by
\bq
E\{\eta(\mathbf{x})|\boldsymbol{\beta}\} = \mathbf{h}(\mathbf{x})^T\boldsymbol{\beta}. \notag
\eq
The vector $\mathbf{h}(\cdot)$ consists of $q$ known regression functions of $\mathbf{x}$, incorporating any beliefs that we might have about the form of $\eta(\cdot)$. In this paper, we use a constant function to illustrate the methodology. The covariance between $\eta(\mathbf{x})$ and $\eta(\mathbf{x'})$ is given by
\bq
\operatorname{cov} \left(\eta(\mathbf{x}),\eta(\mathbf{x'})|\sigma^2,{B},\sigma^2_\epsilon \right)=\sigma^2 c(\mathbf{x},\mathbf{x'}|{B}) + \sigma^2_\epsilon \delta_{\mathbf{x'}\mathbf{x}}, \notag
\eq
conditionally on $\sigma^2$, $\sigma^2_\epsilon$ and ${B}$, where $c(\mathbf{x},\mathbf{x'}|{B})$ is a correlation function that depends on parameters given in ${B}$, $\sigma^2$ is a scaling term for the covariance, $\sigma^2_\epsilon$ is an error or nugget term and $\delta_{\mathbf{x'}\mathbf{x}}$ is a Kronecker delta which is one if and only if $\mathbf{x'}=\mathbf{x}$ and zero otherwise. The function $c(\cdot,\cdot|B)$ must ensure that the covariance matrix of any set of inputs is positive semidefinite. A common choice for this function is the Gaussian  correlation function
\bq
c(\mathbf{x},\mathbf{x'}|{B}) = \exp \left\{ -(\mathbf{x}-\mathbf{x'})^T B(\mathbf{x}-\mathbf{x'})  \right \}, \notag
\eq
where $B$ is a diagonal matrix of roughness parameters. In this paper, we shall simply estimate values of $B$ and $\sigma^2_\epsilon$ using optimisation techniques on the likelihood. 

The output of $\eta(\cdot)$ is observed at $n$ locations, $\mathbf{X}=\{\mathbf{x}_1,\dots,\mathbf{x}_n\}$, to obtain data $\mathbf{y}$. We denote the collection of $n$ observed inputs and outputs as training data  $\mathbf{D}=\bigl((\mathbf{x}_1,y_1),\ldots,(\mathbf{x}_n,y_n)\bigr)^T$. The likelihood of the data given the parameters can be seen in \citet{oh1992} to be 
\begin{equation*}
L(B, \sigma^2_\epsilon; \mathbf{D}) \propto |A|^{-\frac{1}{2}}|H^TA^{-1}H|^{-\frac{1}{2}}(\hat{\sigma}^2)^{\frac{q-n}{2}},
\label{eq: likelihood for b}
\end{equation*}
where 
\begin{align}
H^T &= \left(\mathbf{h}(\mathbf{x}_1)^T,\dots,\mathbf{h}(\mathbf{x}_n)^T\right), \notag \\
A_{ij} &= c(\mathbf{x}_i,\mathbf{x}_j|{B})+ \sigma^2_\epsilon \delta_{ij},\label{eq: mean and cov matrix}\\
{\hat{\sigma}}^2 &= \frac{\mathbf{y}^T\bigl(A^{-1}-A^{-1}H(H^TA^{-1}H)^{-1}H^TA^{-1}\bigr)\mathbf{y}}{n-q-2}.\notag
\end{align}
If we have data in which the output is deterministic, we set $\sigma_\epsilon=0$, so $A_{ij} = c(\mathbf{x}_i,\mathbf{x}_j|{B}).$
We use a weak prior $p(\sigma^2,\boldsymbol{\beta})~\propto~\sigma^{-2}$ for the other hyperparameters as given in \cite{oh1992}. Alternatively,  more informative priors can be used if there are stronger beliefs about the parameters \citep{oak2002}.
 
From properties of the multivariate normal distribution, \citet{oh1992} has shown that
\bq
\left.\frac{\eta(\mathbf{x})-m^*(\mathbf{x})}{\hat{\sigma}\left({\frac{n-q-2}{n-q}c^*(\mathbf{x},\mathbf{x}|B)}\right)^\frac{1}{2}}\right|\mathbf{D},B,\sigma^2_\epsilon \sim t_{n-q}, \notag
\eq
where
\begin{align}
m^*(\mathbf{x}) &= \mathbf{h}(\mathbf{x})^T\hat{\boldsymbol{\beta}} + \mathbf{v}(\mathbf{x})^TA^{-1}(\mathbf{y}-H\hat{\boldsymbol{\beta}}),\notag\\
c^*(\mathbf{x},\mathbf{x}') &= c(\mathbf{x},\mathbf{x'}|{B}) - \mathbf{v}(\mathbf{x})^TA^{-1}\mathbf{v}(\mathbf{x})\notag\\
&+(\mathbf{h}(\mathbf{x})^T\left.-\mathbf{v}(\mathbf{x})^TA^{-1}H)(H^TA^{-1}H)^{-1}(\mathbf{h}(\mathbf{x})^T-\mathbf{v}(\mathbf{x})^TA^{-1}H)^T,\right.\label{eq: post mean and var}\\
\mathbf{v}(\mathbf{x})^T &= \left(c(\mathbf{x},\mathbf{x}_1|{B}),\dots,c(\mathbf{x},\mathbf{x}_n|{B})\right),\notag\\
\hat{\boldsymbol{\beta}} &= (H^TA^{-1}H)^{-1}H^TA^{-1}\mathbf{y}.\notag
\end{align}
Equation (\ref{eq: post mean and var}) shows that the posterior mean, $m^*(\mathbf{x})$, is based on the maximum a posteriori (MAP) estimate of the mean function from the prior, adjusted for on how close the point is to other training points.

\section{PIECEWISE GAUSSIAN PROCESS PRIOR}\label{PGP}
\subsection{Voronoi tessellation with joint centres}\label{vrjc}

\begin{figure}[htb]
\centering
\includegraphics[width=0.9\textwidth]{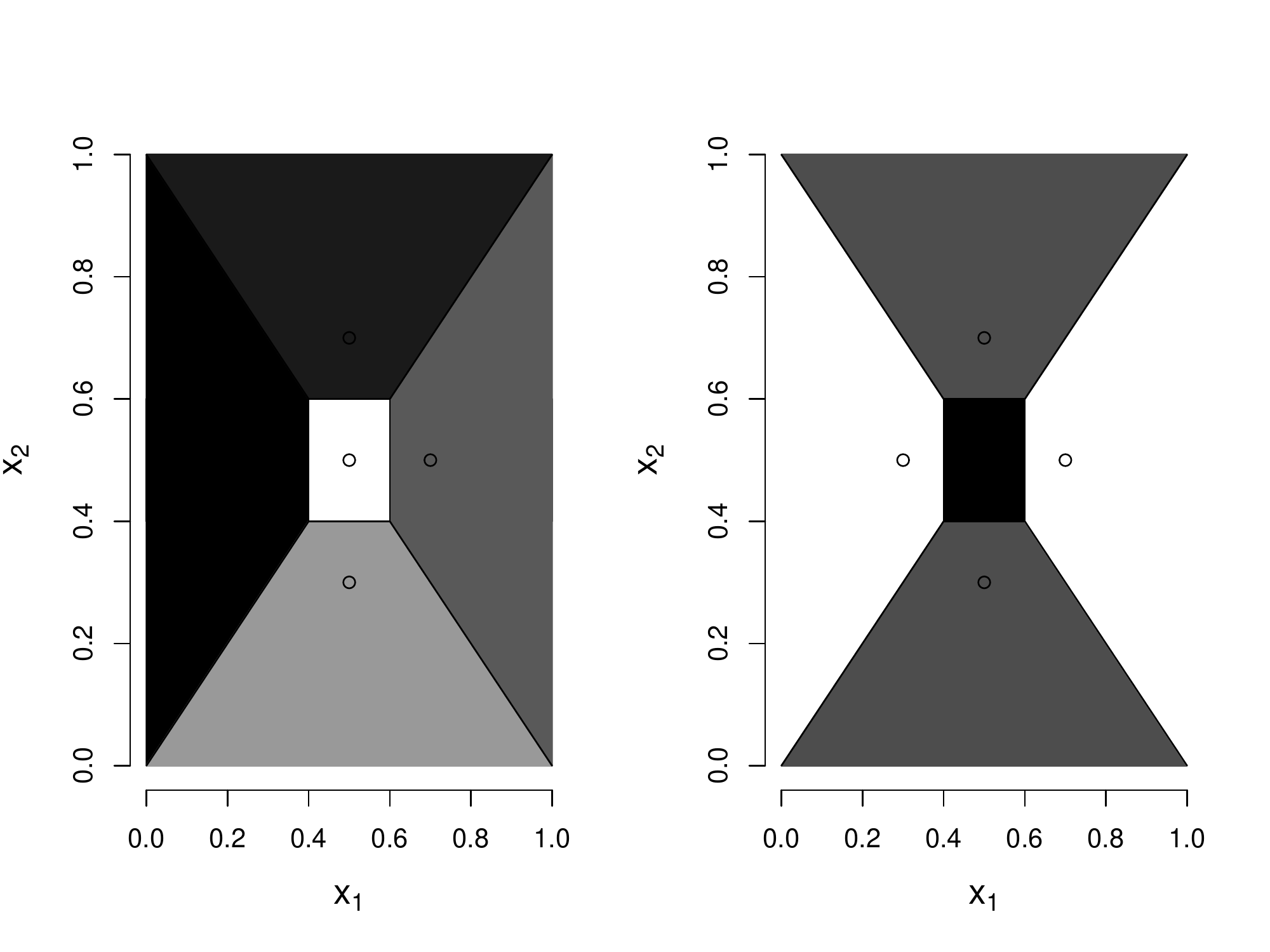}
\caption{Possible regions formed from a Voronoi tessellation, using a simple two dimensional example with five centres.  Regions are indicated by shading tone. Left: All cells are independent, forming five regions with one cell in each region. Right: Cells combine to form three regions.  The left and right cells combine to form one region, as do the top and bottom cells.}
\label{fig:test}
\end{figure}

We allow for discontinuities in the value of $\eta(.)$ by partitioning
the input space $\mathcal{X}$ into $r$ disjoint regions
$R_1, \ldots, R_r$ and denote this partition by
$\mathbf{R}=\{R_1, \ldots,R_r \}$.  The partition structure that we
employ is based on a Voronoi tessellation.  A standard
Voronoi tessellation is defined by a set of $k$ centres,
$\mathbf{x}^\ast_T = \{\mathbf{x}^\ast_{T_1}, \ldots,
\mathbf{x}^\ast_{T_k}\}$.
An arbitrary point $\mathbf{x} \in \mathcal{X}$ is contained in the
cell of the $i$th centre $\mathbf{x}^\ast_{T_i}$ if \bq
d(\mathbf{x},\mathbf{x}^\ast_{T_i})<d(\mathbf{x},\mathbf{x}^\ast_{T_j})~\forall~j
\in \{1,\dots,k \}\backslash i , \notag \eq where
$d(\mathbf{x},\mathbf{x}^\ast_{T_j})$ is the Euclidean distance
between $\mathbf{x}$ and $\mathbf{x}^\ast_{T_j}$.
If we have a finite number of unique, disjoint centres in
finite-dimensional Euclidean space, all of the Voronoi cells are
convex polytopes \citep{Gallier1}.  

To allow for more flexibility than a standard Voronoi tessellation, each
region $R_i$ consists of one or more Voronoi cells. We do not require
Voronoi cells to be neighbours to be in the same region. Hence a
region $R_i$ need not be contiguous or convex, unlike the underlying
Voronoi cells.  Also, note that we do not restrict the centres to be
the training points $\mathbf{x}_i$ 

A simple example of a possible tessellation of $\mathcal{X} = [0,1]^2$ can be seen in Figure~\ref{fig:test}.  On the left is a Voronoi tessellation of five cells.  On the right, the top and bottom cells have joined to form one region, as have the left and right cells; cells in the same region have the same shading. 

The proposed tessellation differs from that used in the PGP method of \citet{kim1} in two important ways.  \citet{kim1} require the centres to be at data points; deciding which data points to used as centres is a key part of fitting a PGP model.  Moreover, all cells are treated independently in the sense that a separate Gaussian process is built in each cell. 
To see why using only Voronoi cells centred on the data points is limiting, consider  an input space that is made up of two different functions on two regions which form a partition of the input space.  We would ideally want to model this using two separate Gaussian processes, one on each region. This can only be modelled accurately using the PGP method if the two regions divide the input space in a way that can be modelled using just two Voronoi cells, each centred on one of the existing data points.  

We could still attempt to model a setup such as this using the PGP
method with a larger number of independent centres.  However, there
are two clear drawbacks to doing so. First, by splitting a region into
multiple independent cells, we are not using all of the points from
that region to estimate the parameters of the Gaussian process, namely
$\boldsymbol{\beta},B,\sigma^2_\epsilon$ and $\sigma^2$. By using a
single region, as allowed for in the proposed approach, all points
from the region can be utilized simultaneously to gain better
parameter estimates. Secondly, using a weak prior distribution for the
GP parameters has the constraint that we need at least four points to
build a Gaussian process with a defined variance, which could make
accurately modelling a function with a discontinuity impossible. We
may, for example, only have five points sampled in a given region and
may not be able to model this region with one centre, making it
inadvisable to split this into multiple regions.

However, this flexibility does come at a price; there are potential
identifiability issues in the proposed approach due to the flexibility
of defining the regions.  A single region which consists of multiple
cells joined together can be equivalent to a region in another model
consisting of a single cell.  This could be easily addressed by
putting a proper prior distribution on the number of cells, $k$, since
the two models will have different number of underlying Voronoi
cells.

We refer to the partition of the $k$ Voronoi cells into $r$ regions as a set of \emph{relationships} between the cells.  In Figure~\ref{fig:test} above, the relationship is that the top and bottom cells form one region, the left and right cells form a second region, and lastly the centre cell forms a region.  Define $\mathcal{C}$ to be a  space such that each element of $\mathcal{C}$ is one of the possible relationships between the $k$ centres, then $c \in \mathcal{C}$ is an index of which relationship is used.

We denote the collection of tessellation parameters by $\mathbf{t}=\{\mathbf{x}_T^\ast, k, r, c\}$ and assign the priors
\begin{align}
\pi(\mathbf{t})&=\pi(k,\mathbf{x}^\ast_T)\pi(r|k)\pi(c|k) \notag ,\\
k,\mathbf{x}^\ast_T &\sim \operatorname{PoiPr}(\lambda) \notag ,\\
r|k &\sim DU(1,k) \notag ,\\
c|k & \sim DU \left(1,b_k\right)  \notag
\end{align}
to model the partition of the input space.  Here $\operatorname{PoiPr}(\lambda)$ is a Poisson process with intensity parameter $\lambda$, $DU(1,k)$ is a discrete uniform distribution on $\{1,\ldots,k \}$, and the $k$th Bell number $b_k$ is the number of all possible ordered partitions \citep{aigner1}.  It should be noted that $\lambda$ is the only hyper-parameter that needs choosing.  Typically, $\lambda$ is chosen by trialling different values until a suitable model is found; alternatively, we could place a prior distribution on it.   There are many adjustments that could be made to incorporate prior beliefs about the underlying model. For example, one adjustment that could be made if appropriate is to replace the Poisson process by one that includes a repulsion term, such as a Gibbs process \citep{illian1}. Using a repulsion term would have the benefit of additional centres having a localised effect on the model tessellation, but in all examples we have studied, we have found the Poisson process to work well.

Combining the separate Gaussian processes on the different regions
yields the likelihood  \bq
L(\mathbf{t},\mathbf{B},\boldsymbol{\beta},\mathbf{\sigma}^2,\mathbf{\sigma}^2_\epsilon;\mathbf{D})\propto\prod^r_{i=1}f_i(\mathbf{y}_i|\mathbf{x}_i,B_i,\sigma^2_i,\sigma^2_\epsilon,\mathbf{\beta}_i,\mathbf{t}),
\notag \eq where
$f_i(\mathbf{y}_i|\mathbf{x}_i,B_i,\sigma^2_i,\sigma^2_\epsilon,\mathbf{\beta}_i,\mathbf{t})$
is the multivariate Gaussian distribution for outputs $\mathbf{y}_i$
corresponding to inputs $\mathbf{x}_i$ which lie in the $i$th
region. We can analytically integrate over $\boldsymbol{\beta}$ and
$\boldsymbol{\sigma}^2$ to give the posterior distribution

\begin{equation}
\pi(\mathbf{B},\mathbf{t},\mathbf{\sigma}^2_\epsilon|\mathbf{D})\propto \prod^r_{i=1}|H^T_iA^{-1}_iH_i|^{-\frac{1}{2}} |A_i|^{-\frac{1}{2}}\Gamma\left(\frac{n_i-q}{2}\right)\left[\frac{2}{(n_i-q-2)\hat{\sigma}^2_i}\right]^{\frac{(n_i-q)}{2}}, \notag
\label{eq: integrated like} 
\end{equation}
where $n_i$ is the number of data points in the $i$th region, $\Gamma( \cdot )$ is the gamma function and $H_i$, $A_i$ and $\hat{\sigma}^2_i$ are as defined in (\ref{eq: mean and cov matrix}), with the subscript $i$ showing that these terms are evaluated using the points that lie in the $i$th region. 

The posterior distribution for $\mathbf{B}$ and $\sigma^2_\epsilon$ is analytically intractable, and we select the parameter values by maximising the likelihood of the $B_i$ for each region and $\sigma^2_\epsilon$. Alternatively, we could take uncertainty in these parameters into account by placing proper prior distributions on them and including them within the MCMC method; however, this has been found to have minimal impact on the resulting predictions and associated uncertainty \citep[for example, see][]{abt1,nagy1}.

\subsection{MCMC implementation}

\noindent
We use reversible-jump MCMC (RJMCMC) \citep{Green1} to estimate the posterior distributions of the model parameters. We require different move types to accommodate the model elements to update: the set of centres for the tessellation and the relationship between the centres. To update the set of centres, we add, take away, or move a centre: these moves are called \textit{birth}, \textit{death} and \textit{move} respectively. To update the relationship between the centres we change a single centre to be in a different region, possibly a new region with no other centre; this move is called \textit{change}. This gives us four possible general moves.

\begin{algorithm}                      
\caption{The RJMCMC implementation of the Joint centre Voronoi Gaussian process}          
\label{alg: MCMC}                           
\begin{algorithmic}                    
    \STATE Begin with a random valid tessellation $\mathbf{t}_0$;
    \FOR{$i=1,\dots,n_s$}
    \STATE Propose \textit{Birth}, \textit{Death}, \textit{Move} or \textit{Change} with equal probability;
    \IF{\textit{Birth} proposed}
        \STATE $\mathbf{x}^\ast_{T(k+1)}=P; P\sim \operatorname{U}_k(\mathbf{0},\mathbf{1})$;
        \STATE Update $c$ such that $\mathbf{x}^\ast_{T(k+1)}$ is combined with an existing region or forms a new region;
    \ELSIF{\textit{Death} proposed}
        \STATE Remove one element of $\mathbf{x}^\ast_{T}$ at random;
        \STATE Remove the chosen point's relationship in $c$;
    \ELSIF{\textit{Move} proposed}
        \STATE Select an element of $\mathbf{x}^\ast_{T}$ at random, $\mathbf{x}^\ast_{T_j}$ say;
        \STATE Propose a new centre $P \sim N_k(\mathbf{x}^\ast_{T_j},\Sigma_p)$, with $\Sigma_p$ tuned for mixing;
        \STATE Set $\mathbf{x}^\ast_{T_j}=P$;
    \ELSIF{\textit{Change} proposed}
        \STATE Select an element of $\mathbf{x}^\ast_{T}$ at random, $\mathbf{x}^\ast_{T_j}$ say;
        \STATE Change $\mathbf{x}^\ast_{T_j}$'s relationship in $c$ s.t. it is independent or related to a different region;
    \ENDIF
    \STATE Update the current tessellation $\mathbf{t}_i$ to obtain proposed tessellation $\mathbf{t}_p$;
    \STATE Fit an independent Gaussian process to each region in $\mathbf{t}_p$;
    \STATE Choose $B$ and $\sigma^2_\epsilon$ to be the values that maximise the likelihood~(\ref{eq: likelihood for b});
    \STATE Calculate the posterior of the proposed model $\pi(\mathbf{B},\mathbf{t}_p,\sigma^2_\epsilon|\mathbf{D})$;
    \STATE Generate $U_i \sim U[0,1]$;
    \IF{$U_i\leq \frac{\pi(\mathbf{B},\mathbf{t}_p,\sigma^2_\epsilon|\mathbf{D})}{\pi(\mathbf{B},\mathbf{t}_i,\sigma^2_\epsilon|\mathbf{D})}\times\text{Prior ratio}\times\operatorname{Adjustment}$}
    \STATE $\mathbf{t}_{i}=\mathbf{t}_p$; 
	\ELSE
	\STATE $\mathbf{t}_{i}=\mathbf{t}_{i-1}$; 
	\ENDIF 
    \ENDFOR
\end{algorithmic}
\end{algorithm}

These four types of proposal are taken to be equally likely during the proposal step. We use an acceptance ratio that is the same as that described in \citet{Green1}, which has the form $$\alpha =\min \left(1,\operatorname{Likelihood~ratio}\times \operatorname{Prior~ratio} \times \operatorname{Proposal~ratio} \right).$$

Due to the setup of the moves, we find that the acceptance ratio simplifies to the ratio of the likelihood of the proposed model to that of the existing model. As we cannot have a death when we have one centre and we also cannot change the relationship of the centre, we only propose birth and move steps in that situation. To maintain reversibility here, when a birth step is proposed, we multiply the acceptance ratio by $1/2$, and, conversely, we multiply the acceptance ratio by 2 when we have two centres and we propose a death step. These multiplications are called $\operatorname{adjustments}$, and we set the $\operatorname{adjustment}$ to 1 in all other cases. Pseudo-code detailing the RJMCMC implementation is given in Algorithm \ref{alg: MCMC}.

After the RJMCMC update of the tessellation, we fit independent Gaussian processes to each region. We can then use the Gaussian process model on each region to make predictions at points in the same region.

\section{ADAPTIVE SAMPLING TO IDENTIFY DISCONTINUITIES}\label{SM}

In some applications, it may be possible to gather additional data at new training points $\mathbf{x}^\ast$. In many cases, this is costly and/or time consuming, so these values of $\mathbf{x}^\ast$ must be chosen with care.  Some applications, such as the cloud modelling we discuss in Section~\ref{EG}, have a number of regions separated by discontinuities and one region is of particular interest.  In such cases, learning more about both that region's boundary and the surface close to the boundary can be of great importance.  In particular, we may wish to sample additional design points close to the  boundary such that we estimate any discontinuities or borders between regions more accurately. Having more information around the discontinuity will not only help us predict outcome values at unobserved locations with more accuracy, but will also supplement the understanding we have about where the discontinuities are occurring, which is often of practical interest. Common sampling methods such as space-filling algorithms and largest uncertainty samplers \citep{santner1} are not tailored to this objective, and adaptive samplers for Gaussian processes have other objectives like reducing overall error or reducing uncertainty \citep{williams1,loeppky1}.

We propose the following sampling method to help estimate these boundaries. The approximate MAP model is found by looking at which tessellation in the posterior sample has the largest likelihood value. The method looks at the MAP model and samples points on the boundary of the region of interest in this model, which is taken to be a good estimate of the boundary of the discontinuity.  There are an infinite number of positions that we could sample on the boundary, and so we attempt to maximise the information we get from each sample. We iteratively choose points on this boundary that are furthest from all existing design points to try to attain some of the properties that are established for space-filling designs. Pseudo-code for this sampling method is given in Algorithm~\ref{alg: bound samp}. 

We note that it would be straightforward to extend this sampling method to sample on the boundary of any region or multiple regions. A change could be made to the algorithm if we are able to double the number of points that we can sample. Instead of sampling at a point $\tilde{\mathbf{x}}_j$, we could look at the line that interpolates $\tilde{\mathbf{x}}_j$ and the centre of its corresponding cell $\mathbf{x}_{T_i}$, then sample two points on this line at distances $||\tilde{\mathbf{x}}_j - \mathbf{x}_{T_i}|| \pm \epsilon$ from the centre. That is, rather than sampling at the point $\tilde{\mathbf{x}}_j$, we sample at points $\tilde{\mathbf{x}}_j\pm \epsilon|\mathbf{x}_{T_i}-\tilde{\mathbf{x}}_j|$, where $0 < \epsilon \ll 1$. This adaptation should in theory sample just inside and just outside of  the discontinuity if $\epsilon$ is chosen suitably. It would also be straightforward to adjust this sampling approach to generate points on or near boundaries giving preference to points with higher posterior uncertainty if we wished to improve both boundary detection and function estimates.

\begin{algorithm}                      
\caption{The boundary sampling method}          
\label{alg: bound samp}                    
\begin{algorithmic}  
    \REQUIRE $n_p>0$ --- number of points to sample;                   
    \REQUIRE $\mathbf{X}$ --- the $n$ locations with observed data;
    \REQUIRE $\mathbf{y}$ --- the $n$ outputs corresponding to locations $\mathbf{X}$;
    \STATE Implement the method from Section \ref{vrjc} to gain a posterior sample of models;
    \STATE Find the MAP model from these posterior samples;
    \STATE Identify the region whose boundary is to be investigated;
   \STATE Randomly sample a candidate set of $n^\ast \gg n_p$ points $\tilde{\mathbf{X}}$ on the boundary of the region; 
   \FOR{$k=1,\dots,n_p$}
	    \STATE Select the point $\tilde{\mathbf{x}}_i$ such that 
            \begin{equation*}
              i = \argmax_{i \in 1,\dots,n^\ast} \min_{j \in 1,\dots,n} d(\tilde{\mathbf{x}}_i,\mathbf{x}_{j});
            \end{equation*}
	    \STATE Remove $\tilde{\mathbf{x}}_i$ from $\tilde{\mathbf{X}}$ and add it to $\mathbf{X}$;
            \STATE Update $n \leftarrow n+1$;
    \ENDFOR
	\STATE Sample outputs $\eta(\cdot)$ at locations $\{\mathbf{x}_{n+1}, \ldots, \mathbf{x}_{n+n_p} \}$.
\end{algorithmic}
\end{algorithm}

\section{ILLUSTRATIVE EXAMPLES}\label{TOY}
\subsection{A diamond-shaped discontinuity}\label{Diamond shaped disc}

\begin{figure}[ht]
\centering
\includegraphics[width=0.8\textwidth]{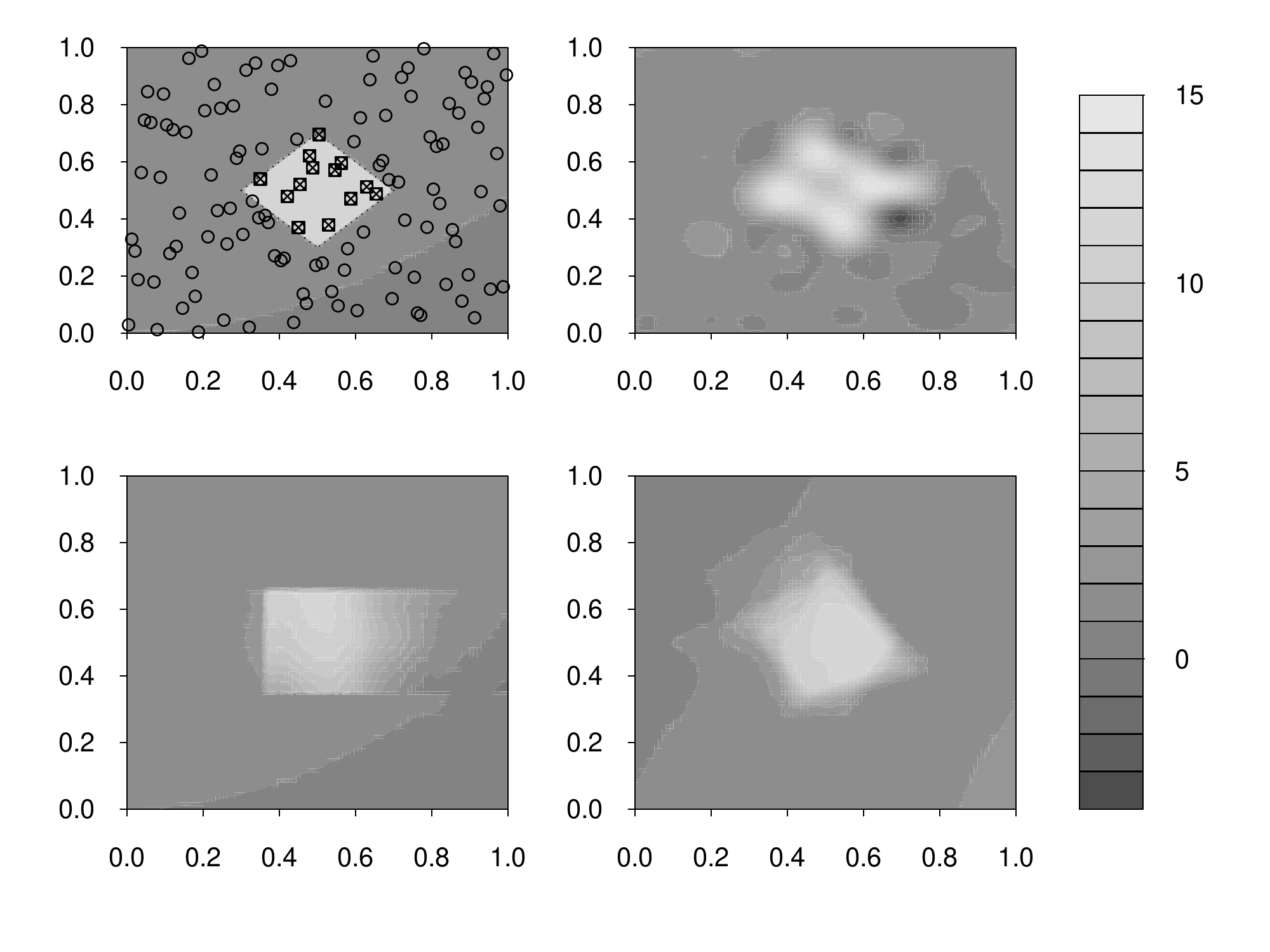}
\caption{Top left: The true diamond test function and design points used; design points  within the discontinuity are square and those outside are shown as circles.  Top right: The standard Gaussian process mean surface.  Bottom left: The TGP integrated surface.  Bottom right: The integrated surface of the proposed method.}
\label{fig: Diamond pictures}
\end{figure}

To initially test the proposed modelling approach, we apply it to the deterministic test function shown in Figure \ref{fig: Diamond pictures}. The example has a discontinuity defined by straight lines, but these are not parallel to the parameter axes . The function is defined by
\begin{equation*}
\eta_1(\mathbf{x})=\begin{cases}
\sin(x_1)+\cos(x_2) \qquad \qquad ~ &\text{for } \mathbf{x} \in T ,\\
\sin(x_1)+\cos(x_2)+10 \qquad &\text{else},
\end{cases}
\label{eq: diamond test fun}
\end{equation*}
where $T=\{\mathbf{x}: x_2-x_1\leq 0.2\, \cap\, x_2-x_1\geq -0.2\, \cap\, x_2+x_1\geq 0.8\, \cap\, x_2+x_1\leq 1.2  \}.$

We evaluate the function at 80 design points, chosen using a Latin hypercube design with a maximin criterion to get a good coverage of the input space \citep{Johnson1}. One thing of interest here is to inspect the surface estimate produced by the proposed model and how this compares to the true surface. Due to the nature of the posterior samples, any estimated surface obtained from a single sample would be conditional on the tessellation $\mathbf{t}_i$ for that sample. We can numerically integrate over $\mathbf{t}$ via Monte Carlo methods using the posterior sample and, hence, have an integrated mean surface that is not conditional on the tessellation parameter; that is an estimate of $E_{\mathbf{t}}\left(E_{\eta(.)}(.)\right)$. This will be referred to as the \textit{integrated surface} whilst the surface of a single sample will be referred to as a \textit{mean surface}. To create the integrated surface, we find the value of the mean surface for each of the posterior tessellation samples at 10,000 points, using an equispaced grid of $100 \times 100$ points,  and find the pointwise mean of these surface over the samples. 

We compare different analysis methods  using the mean squared error (MSE) of the integrated surface for each method. We find that the MSE of the proposed method ($\mse=1.84$) is smaller than that of both the Treed GP ($\mse=1.98$) and the standard GP ($\mse=2.04$). The MSE of the proposed method compared to the others suggests that the new approach is more representative of the true surface. The integrated surfaces of all of the methods can be seen in Figure \ref{fig: Diamond pictures}. We also note that it performs better than the convolution based Gaussian process ($\mse=2.13$). 

\begin{figure}[ht]
\centering
\includegraphics[width=0.8\textwidth]{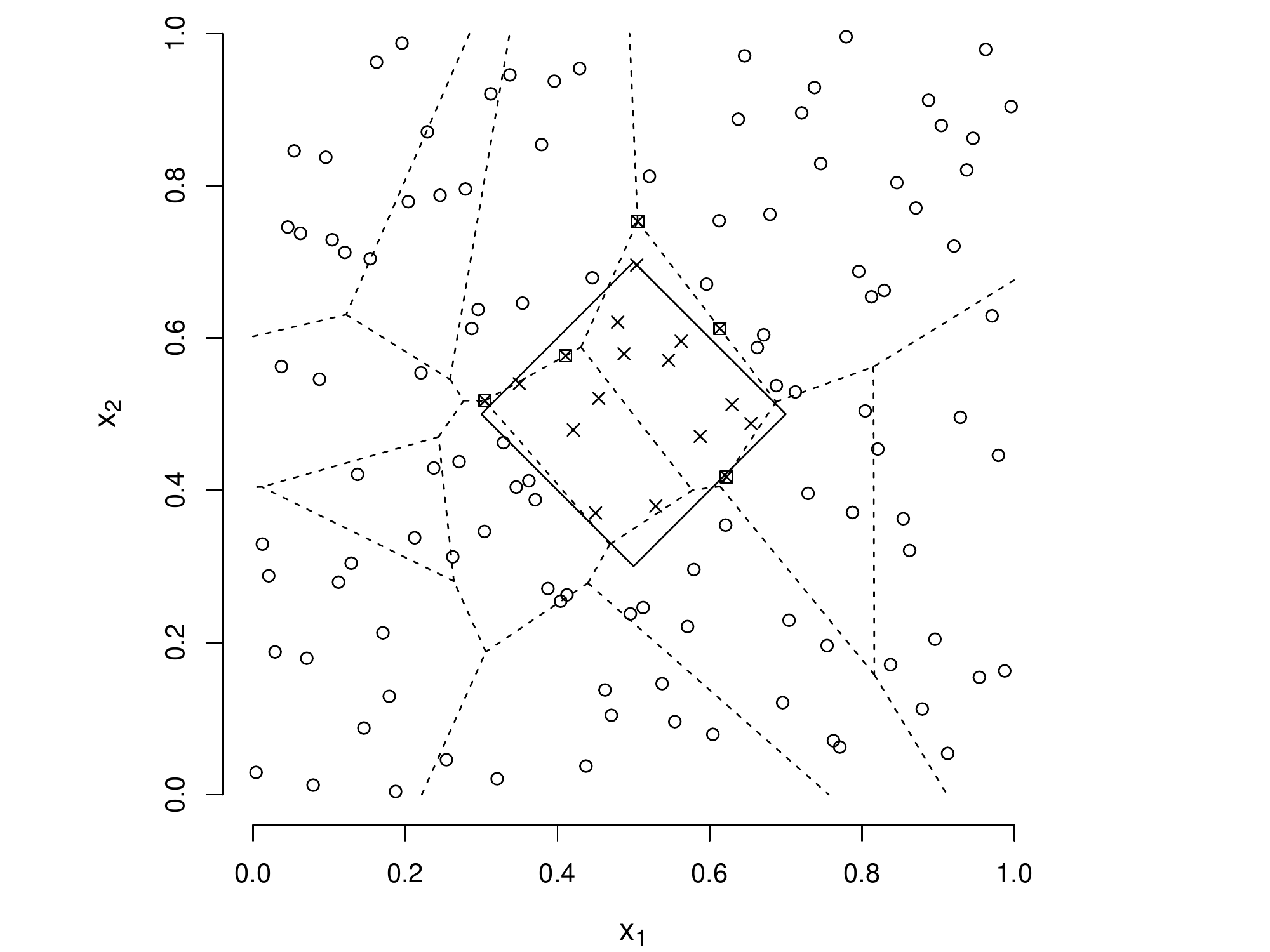}
\caption{MAP tessellation of the diamond example, with the true discontinuity shown as a solid line.  Circles and crosses show  design points  outside and inside the discontinuity respectively.  Cell boundaries from the approximate MAP model are shown as dashed lines. The two cells containing all the discontinuity design points form one region and the remaining cells form a second region.  New design points selected by the adaptive sampler algorithm are denoted by crossed squares.}
\label{fig: MAP and sample diamond 1}
\end{figure}

We also consider the performance of the adaptive sampler for this example. The MAP model from the posterior sample is shown in Figure \ref{fig: MAP and sample diamond 1}. The MAP model has 14 cells divided into two regions, with one region containing 12 of these cells and the other region containing just two. The region with two cells, which contains all of the points from inside the discontinuity, is the region whose boundary we will sample on. To do this, we implement the sampler from Algorithm~\ref{alg: bound samp}, using 2000 candidate points on the boundary and selecting five of these points to evaluate and include in the training data.

\begin{table}[ht]
\centering
\begin{tabular}{cccccccc}\cline{2-8}
& \multicolumn{7}{c}{Number of regions}\\

&1 &2& 3&4&5&6&7\\
\hline
Original points & 0.000 & 0.294 & 0.505 & 0.118 & 0.047 & 0.023 & 0.014\\
After sampler & 0.000 & 0.595 & 0.280 & 0.060 & 0.021 & 0.031 & 0.006\\\hline
\end{tabular}
\caption{Posterior probabilities for the number of regions in the diamond example before and after the adaptive sampler was used.}
\label{tab: results of diamond samp prob}
\end{table}

We can see in Figure \ref{fig: MAP and sample diamond 1} that two of the points we have chosen to sample  lie very close to the true discontinuity, and, around those areas, we should have a much better understanding about the location of the boundary. We will also reduce the uncertainty about the mean function around the other three points that have been sampled although these points do not lie as close to the boundary as the two previously mentioned. We compare the proposed adaptive sampling method to two existing methods of selecting new design points: using a Sobol sequence \citep{giunta1} and selecting the points in $\mathcal{X}$ that have the largest posterior variance. We find that the adaptive sampling algorithm produces the smallest MSE ($\mse=1.352$) using the method from section \ref{SM} when an additional five points are sampled compared to using Sobel ($\mse=1.511$) and the largest posterior uncertainty ($\mse=1.392$).  Table \ref{tab: results of diamond samp prob} reports the posterior probability of $r$ regions for $r = 1, \ldots, 7$ using the proposed method on the original 80 design points and on the 85 points including those chosen by the adaptive sampler.  The most probable number of regions is not equal to the true number of regions when only the original points are considered. However, the true number of regions becomes the most probable when the additional new points are added. In fact, as we further sample more points, we become more confident that the number of regions is two ($\Pr(r=2)=0.89$ when we sample an additional 10 points using the adaptive sampler).

\subsection{A discontinuity with curved boundaries} \label{curved boundaries}

\begin{figure}[th]
\centering
\begin{subfigure}{.5\textwidth}
  \centering
  \includegraphics[width=.9\linewidth]{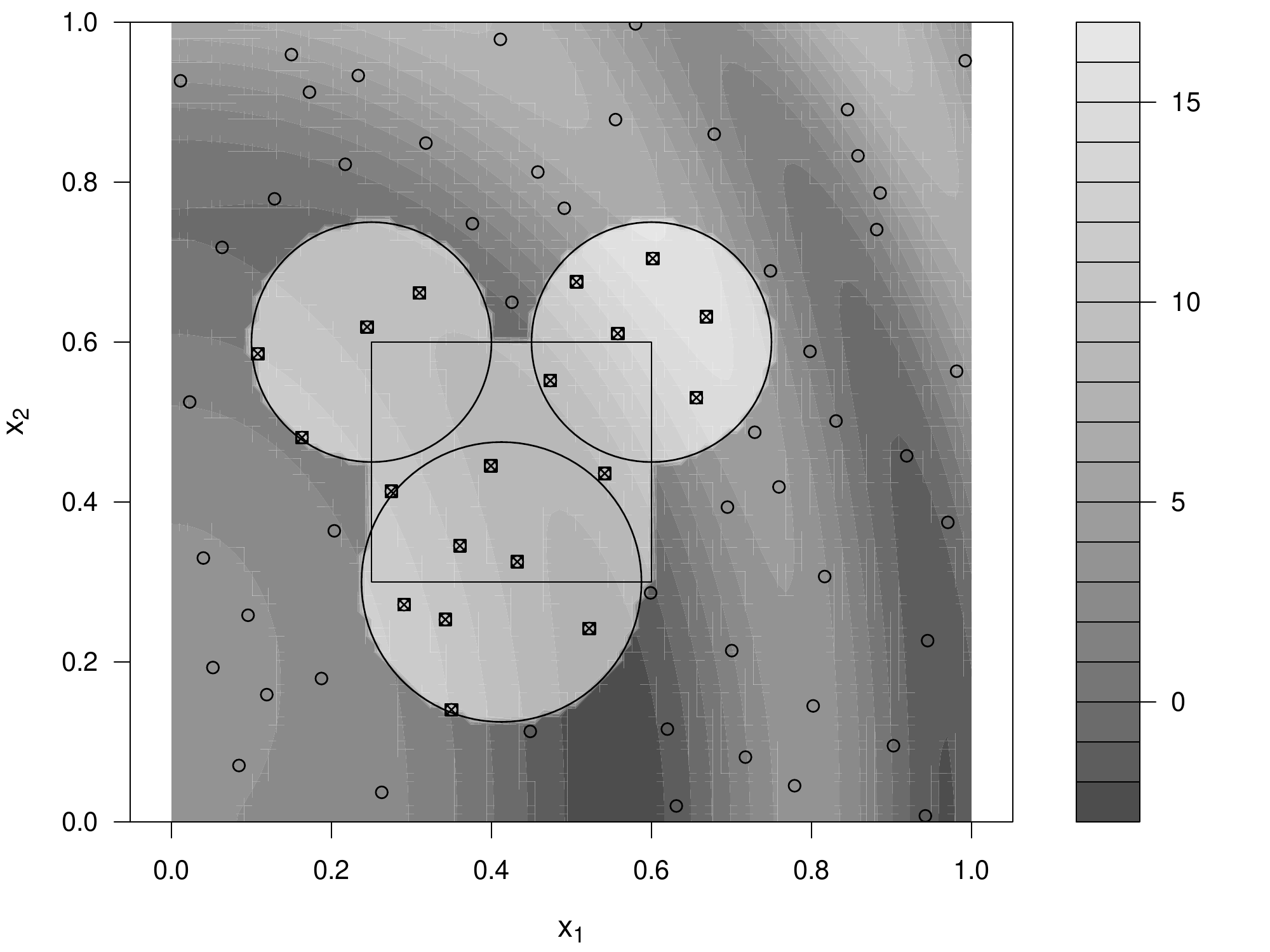}
  \caption{The true function.}
  \label{fig: Mciky true}
\end{subfigure}%
\begin{subfigure}{.5\textwidth}
  \centering
  \includegraphics[width=.9\linewidth]{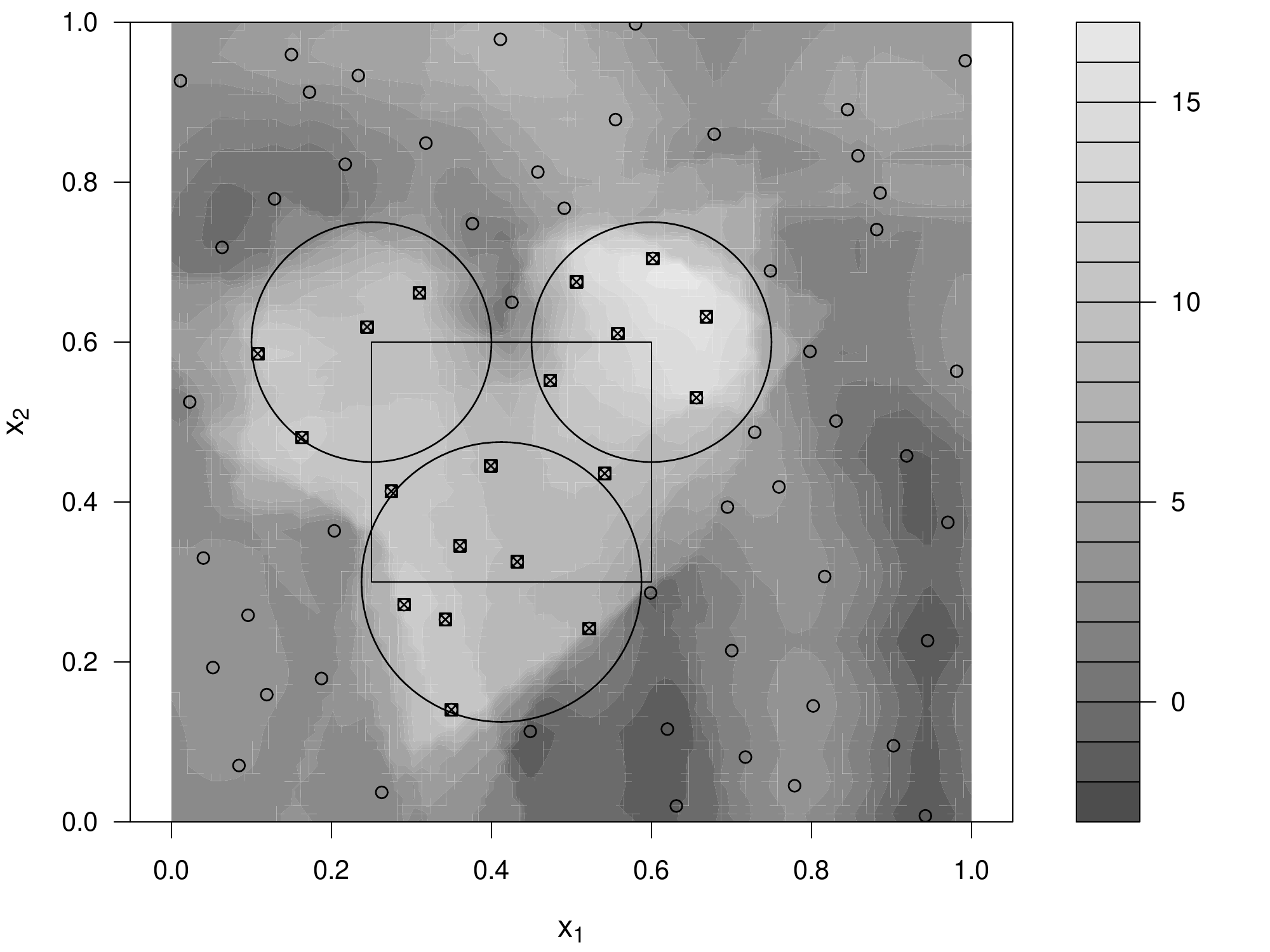}
  \caption{The integrated surface.}
  \label{fig: fig: Micky average}
\end{subfigure}
\caption{Filled contour plots of the true function from equation (\ref{eq: micky mouse}) and the integrated surface of the proposed method.}
\label{fig: micky mouse pics}
\end{figure}

The second test function, shown in Figure \ref{fig: Mciky true}, is a particularly difficult one as precisely representing a circular boundary using Voronoi tessellation would need an infinite number of centres. This test function is defined as
\begin{equation*}
\eta_2(\mathbf{x})=\begin{cases}
x_1^2+5x_2^2+3\cos(10x_1^2+5x_2^2) +10 \qquad \text{for } \mathbf{x} \in L, \\
x_1^2+5x_2^2+3\cos(10x_1^2+5x_2^2) \qquad \qquad ~ \text{else},
\end{cases}
\label{eq: micky mouse}
\end{equation*}
where 
\begin{align*}
L&=\big\{\mathbf{x}: \{x_1 \in [0.25,0.6]\cup  x_2 \in [0.3,0.6]\} \cup \{(x_1-0.25)^2+(x_2-0.6)^2\leq0.15^2 \} \\
& \cup \{ (x_1-0.6)^2+(x_2-0.6)^2\leq0.15^2 \} \cap \{ (x_1-0.4125)^2+(x_2-0.3)^2\leq0.175^2 \} \big\}.
\end{align*}

We evaluate the function at 70 design points chosen using a Latin square design with a maximin criterion to ensure  even coverage of the input space \citep{Johnson1}. The integrated surface obtained by the proposed method can be seen in Figure \ref{fig: fig: Micky average}. We can see that the method has performed as well as can be expected when considering the data we have used to train it. The mean squared error of the integrated surface for the proposed method ($\mse=4.498$) is lower than that for the TGP ($\mse=6.886$) or the standard GP ($\mse=6.473$). 

In the TGP method, the input space is partitioned using non-overlapping straight lines parallel to the parameter axes and a separate Gaussian process is built for each region. We can see from the shape of $L$ in Figure \ref{fig: Mciky true} that we would need a large number of these regions to be able to get a good approximation for the the true shape of the discontinuity. A similar argument follows when we consider the shape of $T$ in the simulated example of Section~\ref{Diamond shaped disc}. Since a standard GP is inappropriate for both of these functions due the smoothness assumption clearly being violated. As a result the mean function must over-smooth to ensure that the function intersects the training points exactly leading to poor estimates around the discontinuity, leading to high MSE values.

\section{APPLICATION: Cloud modelling}\label{EG}

We illustrate the proposed method on simulation data from a complex numerical cloud-resolving model, the System for Atmospheric Modelling (SAM) \citep{Khairoutdinov1}. For this example, the model is used to simulate the development of shallow nocturnal marine stratocumulus clouds for 12 hours over a domain of size 40 km x 40 km x 1.5 km height. Initial conditions are described through a vector $\mathbf x$ of six key parameters. These simulations are an updated version of the nocturnal marine stratocumulus simulations (set 2) described in detail in \citet{f1}, with longer run-time and updated radiation scheme.  We focus here on the average predicted cloud coverage fraction over the domain in the final hour of the simulations, $y$.

Shallow clouds are very important to the climate system as they reflect solar energy to space and hence cool the planet, offsetting some of the greenhouse gas warming. These clouds are particularly sensitive to aerosol concentrations in the atmosphere and meteorological conditions, where small changes in temperature and humidity profiles can  impact strongly on whether clouds form or not, and how thick/reflective they are. It is essential to understand how changes in aerosol and meteorological conditions can affect shallow clouds in order to improve their representation in climate models. Currently, large-scale climate model representations of shallow clouds are poor as they form and develop on smaller scales than the large grids used, yet how they are represented can have a strong influence on predictions of climate sensitivity, in particular the magnitude of warming for a prescribed increase in CO2. 

Initial investigations of these SAM simulations and expert opinion has suggested that the model potentially produces two different forms of cloud behaviour, open and closed cell behaviour, over the six-dimensional parameter space.  Hence, the underlying  function  we wish to model is potentially made up of a single function with two regimes, likely containing a discontinuity in $y$ as the model behaviour moves between these regimes. As such, there is also interest in knowing about the location of any discontinuity/change in regime in order to explore where and why this phenomenon occurs.
 
\begin{figure}[ht]
\centering
\includegraphics[width=\textwidth]{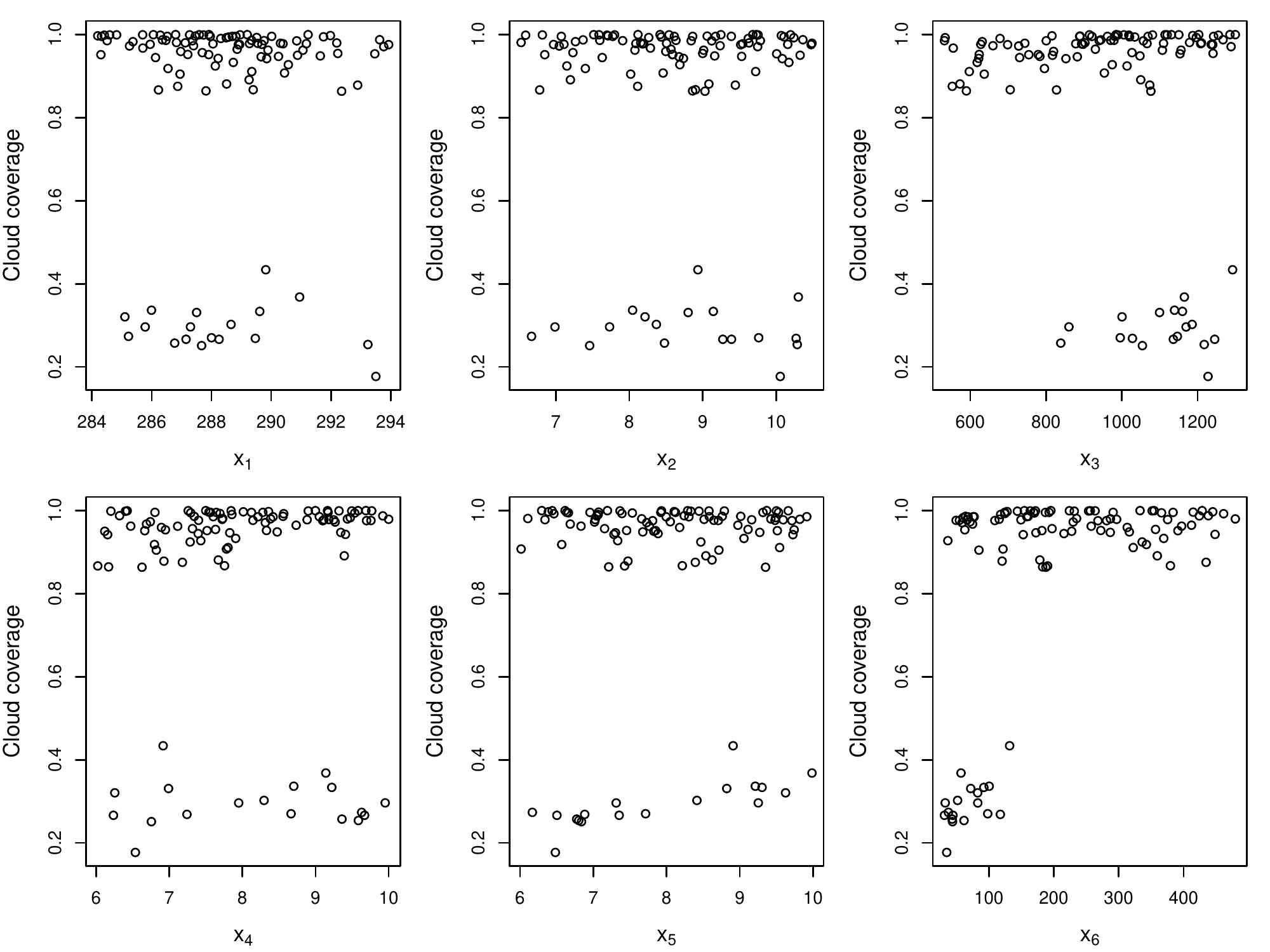}
\caption{The scatterplots of each input plotted against the output for the original 105 cloud coverage data-points.}
\label{fig: input v output cloud}
\end{figure}

We have 105 training points available from the simulations to model the cloud coverage fraction, $y$, where the input combinations  were chosen to cover the 6-d parameter space using a space-filling maximin Latin hypercube design. The six parameters investigated here are listed in Table~\ref{tab:inputs}, and scatter plots of the outputs against the individual inputs can be seen in Figure \ref{fig: input v output cloud}, in which there is clear evidence of two regimes but no obvious way of splitting the data according to a single input parameter. We see from Figure \ref{fig: input v output cloud} that most areas of the parameter space output consistent values of $y$ at around 0.9; however, some other areas have much smaller values of $y$, from 0.2 to 0.4.  The plot of the output against the aerosol concentration ($x_6$) input suggests that high values of this variable are very likely to yield large cloud coverage values; however, there is no clear way to differentiate low values. 

\begin{table}[ht]
\centering
\begin{tabular}{clc}\hline
Label & Input description & Range investigated \\
\hline
$x_1$ & Liquid water potential temperature  & 284--294 K \\
$x_2$ &  Total non-precipitating water mixing ratio &  6.5--10.5 g$\cdot\mbox{kg}^{-1}$\\
$x_3$ & Depth of mixing layer  &  500--1300 m\\
$x_4$ & Jump in water potential temperature at inversion  &  6--10 K\\
$x_5$ &  Jump in water mixing ratio at inversion  &  6--10 g$\cdot\mbox{kg}^{-1}$\\
$x_6$ & Aerosol concentration  &  30--500 $\mbox{cm}^{-3}$\\\hline
\end{tabular}
\caption{Input parameters under investigation and ranges used for each in the analyses \citep[following][]{f1}.}
\label{tab:inputs}
\end{table}

Given that the model output here is bounded between zero and one, some transformation may improve the fitting results. For instance, a logit-transformation might be thought to be better suited to a Gaussian model \citep[like in][]{henderson1,andrianakis1} or a spatial process designed to model proportions directly \citep{paradinas1}. Prior to fitting the partitioning models, we fitted a standard GP model to the logit-transformed output, and we found the GP performed very poorly because the step in the function is exacerbated by the transformation.  We also fitted the TGP and the proposed model to the logit-transformed data and found that they both produced worse estimates than for the raw output fit. Here this can be explained by the outputs being at the high end of the percentage range.

The posterior sample from the proposed methodology indicates that the MAP model obtained contains two regions. The posterior distribution for the number of regions, shown in Table \ref{tab: prob for num regions}, indicates that two regions ($\Pr(r=2|\mathbf{D})=0.667$) is the most probable number, despite the fact that no prior knowledge of this was incorporated. 

\begin{table}[ht]
\centering
\begin{tabular}{cccc}\hline
Number of regions &1 & 2 & 3\\
\hline
Probability &0.102 &0.667 &0.231 \\\hline
\end{tabular}
\caption{The posterior probability distribution for the number of regions.}
\label{tab: prob for num regions}
\end{table}

A further 35 simulations with different input parameter configurations to the training data were run through the computer simulator and used for validation, following the advice of \citet{bastos1}. The proposed method performs better at predicting these validation points ($\mse=0.016$) than the TGP ($\mse=0.032$) and the standard GP ($\mse=0.025$) methods. Following this, we refitted the model using all 140 simulations as training points, and then the adaptive sampler method from Section~\ref{SM} was implemented. An additional 25 parameter combinations were selected, chosen using a candidate set of 170,000 points sampled on the boundary of the smaller (low cloud fraction) region, and these simulations were run.

The new simulations were incorporated in the training data set and the model refitted.  The resulting MAP model has two regions, and the posterior sample yields  $\Pr(r=2|\mathbf{D})=0.87$.  The MAP model has 18 Voronoi cells corresponding to one region and 87 corresponding to the other region. The region with 18 cells corresponds to low cloud fraction output and will be referred to as the smaller region. The region with 87 cells corresponds to high cloud fraction output and will be referred to as the larger region. 

\begin{figure}[th]
\centering
\includegraphics[width=0.9\textwidth]{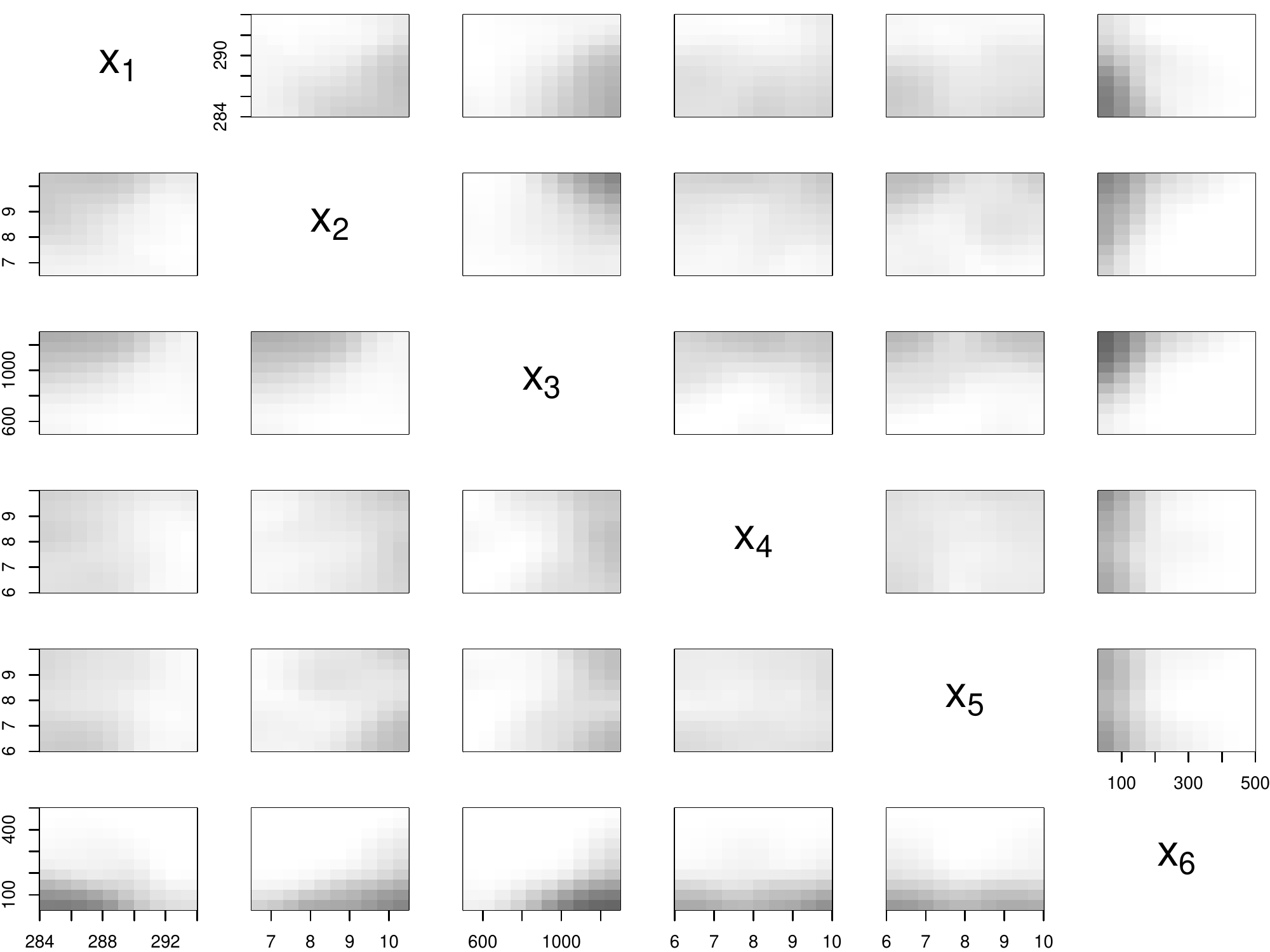}
\caption{The `averaged' proportion of points that lie within the smaller region for each of the 2-d projections based on the MAP model.}
\label{fig: test regions}
\end{figure}

Visualising the shape of the two regions and the discontinuity between them is challenging in $d > 3$ dimensions.  In Figure \ref{fig: test regions}, we attempt to visualise the shape of the boundary of these regions of cloud behaviour. We used ten equispaced points in each input dimension to create a grid of 1,000,000 equispaced points over the six dimensions, and noted which points lie in each region of the MAP model. To aid with visualisation, we perform a dimension reduction technique using a 2-d averaging scheme.  There are 15 possible pairwise combinations of input variables $\left\{(x_1,x_2), (x_1,x_3), \cdots \right\}$, which are each assigned 100 equally spaced points in 2-d. For each of these points, there are 10,000 possible combinations that the other inputs can take and we compute the proportion of these 10,000 points that lie within the smaller region. In Figure \ref{fig: test regions}, we use a grey scale to represent this proportion, with a white  or black block meaning that all of the points lie within the larger or smaller of the two regions respectively, and so correspond to areas of high or low cloud fraction output.

Figure \ref{fig: test regions} shows that the smaller (low cloud fraction) region does indeed appear to have a complex shape in the parameter space as initially suspected. In particular, an interesting aspect of this region can be seen in $x_6$, the aerosol concentration. It appears that smaller values of aerosol concentration are much more likely to be attributed to the smaller region corresponding to low cloud fraction. This observation is supported by the MAP model that was seen when a TGP was attempted with the MAP model splitting the range of the aerosol concentration input variable at 117.7 cm$^{-3}$. Figure \ref{fig: test regions} also indicates a reason why the TGP performed poorly compared to the proposed method; in the $(x_1,x_6),(x_2,x_6)$ and $(x_3,x_6)$ projections, we see that the region appears to have a curved boundary, which the TGP will not be able to model accurately with straight lines.

The results presented here show that, by using the proposed modelling approach, we are able to more clearly and accurately capture and represent the discontinuity that corresponds to the sharp change in cloud behaviour over the six-dimensional parameter space of the cloud model initial conditions.  This is an important result to the cloud modelling community, as this enables the identification of the key initial conditions under which these changes in behaviour may occur. We are also able to determine the sensitivity of the cloud fraction output to the co-varying initial conditions. Full exploration of this may ultimately lead to improvements in the way the shallow cloud coverage is represented in climate models.

\section{Discussion}\label{DIS}

In this paper, we developed a method that can be used to model functions that we believe contain discontinuities or display heterogeneous characteristics. The use of Voronoi tessellations as a tool to partition the input space has been shown to be advantageous over similar methods such as treed GP in the simulations and data we explored. The idea of joining Voronoi tiles is trivial to extend to other methods; for example, we could join partitions of the treed GP to create larger and non-convex regions. 	

There are computational benefits to the proposed approach over fitting a standard Gaussian process model: instead of fitting a Gaussian process model to $n$ function outputs, which requires the inversion of a $n\times n$ matrix and has computing time in $\mathcal{O}(N^3)$, we fit Gaussian process models to sets of outputs that are smaller in size. For instance, if we divide $\mathcal{X}$ into $k$ regions, we have that the largest region could have at most $n-(k-3)$ data points.

Standard Voronoi tessellations were used to partition the input space due to their flexibility. However, using standard Voronoi tessellations still partitions the input space with straight lines and more flexibility over the shape of these partitions would be preferable. One possible extension is the use of weighted Voronoi tessellations. The use of weights on Voronoi tessellations allows for a greater range of partition shapes. For example, we could use multiplicatively weighted Voronoi tessellations to create round partitions or additively weighted Voronoi tessellations to create hyperbolic curves to partition the input space \citep{okabe1}. Another generalization is the additively weighted power diagram or sectional Voronoi tessellation \citep{okabe1}, formed by the intersection of the input space with a Voronoi tessellation in a higher-dimensional space. The cells of the sectional tessellation are again convex polytopes, but the configurations of cells that can occur differ from those in a standard Voronoi tessellation.  Differences can be shown to occur with probability one if the higher-dimensional tessellation is Poisson-Voronoi; see \citet{chiu} for a more precise statement and proof. The use of these weights however will add a new set of parameters to the model that need to be estimated and therefore increase the model complexity. Exploration is needed as to whether the increased flexibility justifies the additional computational cost. Alternatively, we can consider perturbing the individual vertices of the Voronoi cells. Again, this is computationally expensive, but it adds flexibility by allowing polygonal but non-convex cells, as used by \citet{BM2003}.

The methods developed depend on forming partitions of the locations $\mathbf{x}_i$ in order to fit separate Gaussian processes.  There may be some connection to the partitions formed by mixture clustering methods, although the use of the function values $y_i$ in allocating observations to partitions, and the disjoint nature of the regions, is rather different to conventional clustering methods.  

\subsection*{Acknowledgements}

The authors are grateful for the constructive comments of the editors and referees that have led to improvements in this paper.

\bibliography{Master}   

\end{document}